\documentclass{iau} 
\usepackage{graphicx}
\usepackage{natbib}

\pubyear{2022}
\volume{372}
\jname{The Era of Multi-Messenger Solar Physics}
\editors{G.~Cauzzi \& A.~Tritschler, eds.}
\title[Aditya-L1 Mission]{The Aditya-L1 mission of ISRO}
\author[Tripathi et al.]{Durgesh Tripathi$^1$, D. Chakrabarty$^2$, A. Nandi$^3$, B. Raghvendra Prasad$^4$, A. N. Ramaprakash$^1$, Nigar Shaji$^5$, K. Sankarasubramanian$^4$, R. Satheesh Thampi$^6$, V. K. Yadav$^6$}

\affiliation{$^1$Inter-University Centre for Astronomy and Astrophysics, Ganeshkhind, Pune 411007, India \\ [\affilskip]
$^2$Physical Research Laboratory, Ahmedabad 380 009, India \\ [\affilskip]
$^3$ISITE Campus, U R Rao ISRO Satellite Centre, Marathahalli, Bengaluru 560 037, India\\ [\affilskip]
$^4$Indian Institute of Astrophysics, 2nd Block, Koramangala, Bengaluru 560 034, India \\ [\affilskip]
$^5$Old Airport Road, Vimanapura PO
Bengaluru 560017, India\\ [\affilskip]
$^6$Space Physics Laboratory, VSSC, Thiruvananthapuram 695 022, India\\ [\affilskip]}

\begin{document}
\maketitle

\begin{abstract}
The Aditya-L1 is the first space-based solar observatory of the Indian Space Research Organization (ISRO). The spacecraft will carry seven payloads providing uninterrupted observations of the Sun from the first Lagrangian point. Aditya-L1 comprises four remote sensing instruments, {\it viz.} a coronagraph observing in visible and infrared, a full disk imager in Near Ultra-Violet (NUV), and two full-sun integrated spectrometers in soft X-ray and hard X-ray. In addition, there are three instruments for in-situ measurements, including a magnetometer, to study the magnetic field variations during energetic events. Aditya-L1 is truly a mission for multi-messenger solar astronomy from space that will provide comprehensive observations of the Sun across the electromagnetic spectrum and in-situ measurements in a broad range of energy, including magnetic field measurements at L1. 

\end{abstract}

              
\section{Introduction}
The Sun is the life-giving star that has been studied for centuries. In the last 500 years or so, more detailed observations of the Sun's structure -- both internal and atmosphere-- have been studied. After discovering that the outer solar atmosphere was more than a million degrees hot in the 1940s, it was realized that to get a comprehensive understanding of the Sun, it is mandatory to go to space. 

In the last several decades, several spacecraft have been flown by The National Aeronautics and Space Administration (NASA), the European Space Agency (ESA), and the Japan Aerospace Exploration Agency (JAXA). Some of the most important missions are Yohkoh \citep[][]{yohkoh}, the Solar and Heliospheric Observatory \citep[SoHO;][]{soho}, the Transition Region and Coronal Explorer \citep[TRACE;][]{trace}, the Solar Terrestrial Relations Observatory \cite[STEREO;][]{stereo}, Hinode \cite[][]{hinode}, the Solar Dynamics Observatory \cite[SDO;][]{sdo}, the Interface Region Imaging Spectrometer \citep[IRIS;][]{iris} and the two most recently launched missions namely the Parker Solar Probe \citep[PSP;][]{psp} and the Solar Orbiter \citep{so2,so1}. These, in addition to multiple ground-based observatories, have provided a wealth of information that has enhanced our understanding of the working of our star.

\begin{figure}[h]
\begin{center}
  \includegraphics[width=0.72\textwidth]{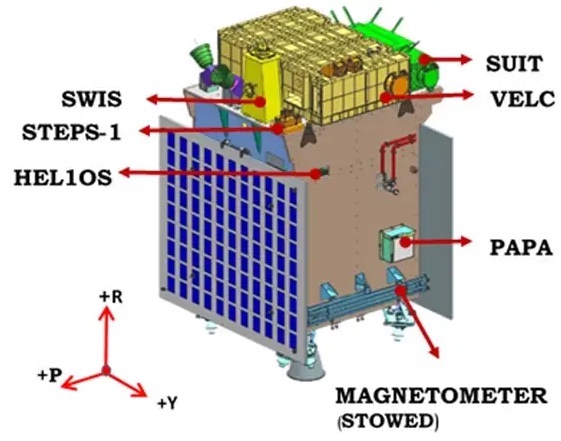} 
  \includegraphics[width=0.55\textwidth]{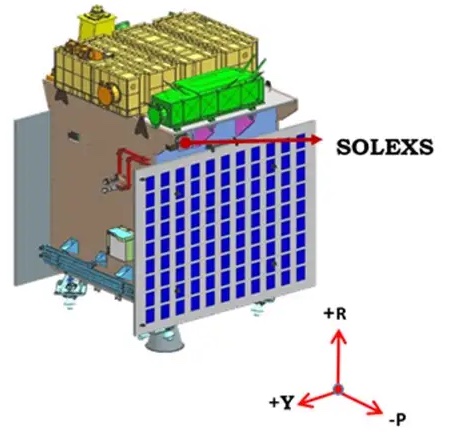} 
 \caption{The model spacecraft indicating all the different payloads onboard Aditya-L1. Image courtesy of Indian Space Research Organization (ISRO)(https://www.isro.gov.in/Aditya\_L1.html).} \label{aditya}
\end{center}
\end{figure}
Although observations and modeling have allowed a great leap toward understanding the Sun, numerous unsolved problems remain. These problems relate to the coupling of the magnetized solar atmosphere, heating of the upper solar atmosphere, nature of solar wind, and dynamics of the inner heliosphere. Moreover, how and why the high energy radiation, particularly in the near ultraviolet, changes and affects the Earth's atmosphere is not fully comprehended. 

One of the most challenging tasks related to coronal heating is measuring the coronal magnetic field. Currently, there is no direct measurement available. Such measurements will not only help understand the heating mechanism but also provide crucial information related to solar eruptions and possibly forecasting the direction of interplanetary magnetic field (IMF) associated with interplanetary coronal mass ejections (ICMEs). The other equally challenging task has been to determine the kinematic profiles of Coronal Mass Ejections (CMEs) during their early phases of evolution. Currently, regular kinematic profiles of CMEs are available using the coronagraphic observations made using the Large Angle and Spectrometric Coronagraph (C2 ad C3) on board SoHO. However, these are only available beyond 2.5~R${\odot}$. However, observations have shown that 90\% of the acceleration in CMEs occur below 2~R${\odot}$ \citep{cme_acc}. Therefore, with the existing facilities, it is impossible to obtain a good handle on the early phase of the kinematics of CMEs. In turn, that becomes one of the primary sources of uncertainties in predicting the arrival time of CMEs. It has also been shown that a significant amount of activity, such as magnetic reconnections leading to the bifurcation of flux rope as well post CME reconnection, may occur well below 2.5~R${\odot}$ \citep[][]{Gilbert, TriS_1, Tris_2}.

CMEs are often associated with flares, which are the intense eruption of electromagnetic radiation. Flares are observed routinely in Soft X-rays using Geostationary Operational Environmental Satellite Network. However, it has been shown that they have counterparts across the electromagnetic spectrum \citep{benz}. While flares are specular in X-ray and have been used for monitoring the space weather, it has been shown that more than $\sim$70\% of the flare energy is contained in other wavelengths such as visible and near-ultraviolet \citep[][]{neidig, Kre_1}. While this may have substantial implications on the physics of solar flares, it may have significant effects on the total and spectral irradiance of the Sun \citep[][]{Kre_2}.

It is also known that the total solar irradiance varies slowly on decadal and longer time scales. The variation during the recent solar magnetic cycle has been about 0.1\% \footnote{Physikalisch-Meteorologisches Observatorium Davos and World Radiation Center (PMOD/WRC)}. However, it is known that solar irradiance shows a strong wavelength dependence and that up to 60\% of the total irradiance variations are produced at wavelength below 400~nm \citep{KriSF_2006}. Of these, the wavelength band of 200{--}400~nm is of particular importance due to its direct effects on the chemistry of Ozone and Oxygen in the stratosphere of Earth. It thereby plays a crucial role in Sun-climate relations. 

Using the observations taken by SOlar Stellar Irradiance Comparison Experiment( SOLSTICE) and Spectral Irradiance Monitor (SIM), both on board SOlar Radiation \& Climate Experiment (SORCE), \cite{HaiW} showed that there is a considerable discrepancy between the solar spectral irradiance observed and modeled by \cite{Lean}. For the wavelength regions below 400~nm, the models under predicts the irradiance, while for the wavelengths larger than 400~nm, the irradiance is over-predicted. Such discrepancy may have a substantial impact on modeling the Sun-climate relations. Note that the observations are recorded using the Sun-as-a-start spectrum. These results suggest that some missing physics is crucial for accurately modeling the sun-climate relations. 

The Aditya-L1 mission is the flagship mission of the Indian Space Research Organization (ISRO). It is ISRO's first solar observatory in space. The satellite will be launched in 2023 by PSLV-XL and inserted in a halo orbit around the First Lagrangian Point (L1), which is about 1.5 million kilometers from the Earth on the Sun-Earth line. The primary science goals of Aditya-L1 mission are:

\begin{itemize}
    \item Diagnostics of the large and small scale structures in the solar corona
    \item Magnetic topology and field measurements in the solar corona
    \item Origin and dynamics of solar flares and coronal mass ejections (CMEs)
    \item Spectral energy distribution in solar flares
    \item Spatially resolved solar spectral irradiance
    \item Formation and dynamics of solar prominences and filaments
    \item Coronal abundances and first ionization potential (FIP) effects
    \item Solar wind composition and particle energy distribution
    \item Measurements of the magnitude and variability of the interplanetary magnetic field (IMF)
\end{itemize}

To address the above-mentioned questions, the spacecraft will carry seven payloads performing remote sensing from Hard X-ray to infrared and in situ measurements in a broad energy range, including interplanetary magnetic field measurements. The seven payloads are:
\begin{itemize}
    \item The Solar Ultraviolet Imaging Telescope (SUIT),
    \item The Visible Emission Line Coronagraph (VELC),
    \item The SOlar Low Energy Spectrometer (SoLEXS),
    \item The High Energy L1 Orbiting X-ray Spectrometer (HEL1OS),
    \item The Aditya Solar Wind Particle EXperiment (ASPEX),
    \item The Plasma Analyser Package for Aditya (PAPA), and
    \item The Magnetometer (MAG)
\end{itemize}

In the rest of the paper, we shall discuss and describe the uniqueness and salient features of different payloads and the science questions they aim to address.
\section{Payloads}
The Aditya-L1 spacecraft carries seven payloads {--} four for remote sensing and three for in-situ measurements. Below we describe each of the payloads.
\subsection{The Solar Ultraviolet Imaging Telescope (SUIT)}
The Solar Ultraviolet Imaging Telescope \citep[SUIT;][]{suit, suit1, suit2} is one of the four remote sensing payloads on board Aditya-L1. It is a combined medium and narrow band filter imager that will observe the Sun in the Near UltraViolet (NUV) wavelength band 200{--}400~nm. With this wavelength range, SUIT aims to cover the lower and middle atmospheric layer of the Sun, i.e., the photosphere and the chromosphere. These two layers for the most crucial for addressing questions related to the dynamic coupling of the magnetized solar atmosphere and various fundamental processes occurring in the partially ionized plasma. Moreover, solar radiation in this wavelength range is fundamental to studying Sun-Climate relations. Therefore, SUIT will make unique measurements of spatially resolved Solar Spectral Irradiance (SSI) in the NUV. Such measurements are crucial for comprehending the chemistry of Ozone and Oxygen in the stratosphere of the Earth and, therefore, central to the Sun-climate relationship.

The primary science questions that SUIT will address are:
\begin{itemize}
    \item the magnetic coupling between the lower and the middle solar atmosphere,
    \item the spatially resolved SSI in the NUV (200{--}400~nm),
    \item the energetics of solar flares in the photosphere and chromosphere,
    \item the dynamics of eruptive and non-eruptive prominences and filaments and
    \item to provide the context to the VELC.
\end{itemize}

\begin{figure}[h]
\begin{center}
 \includegraphics[width=0.6\textwidth]{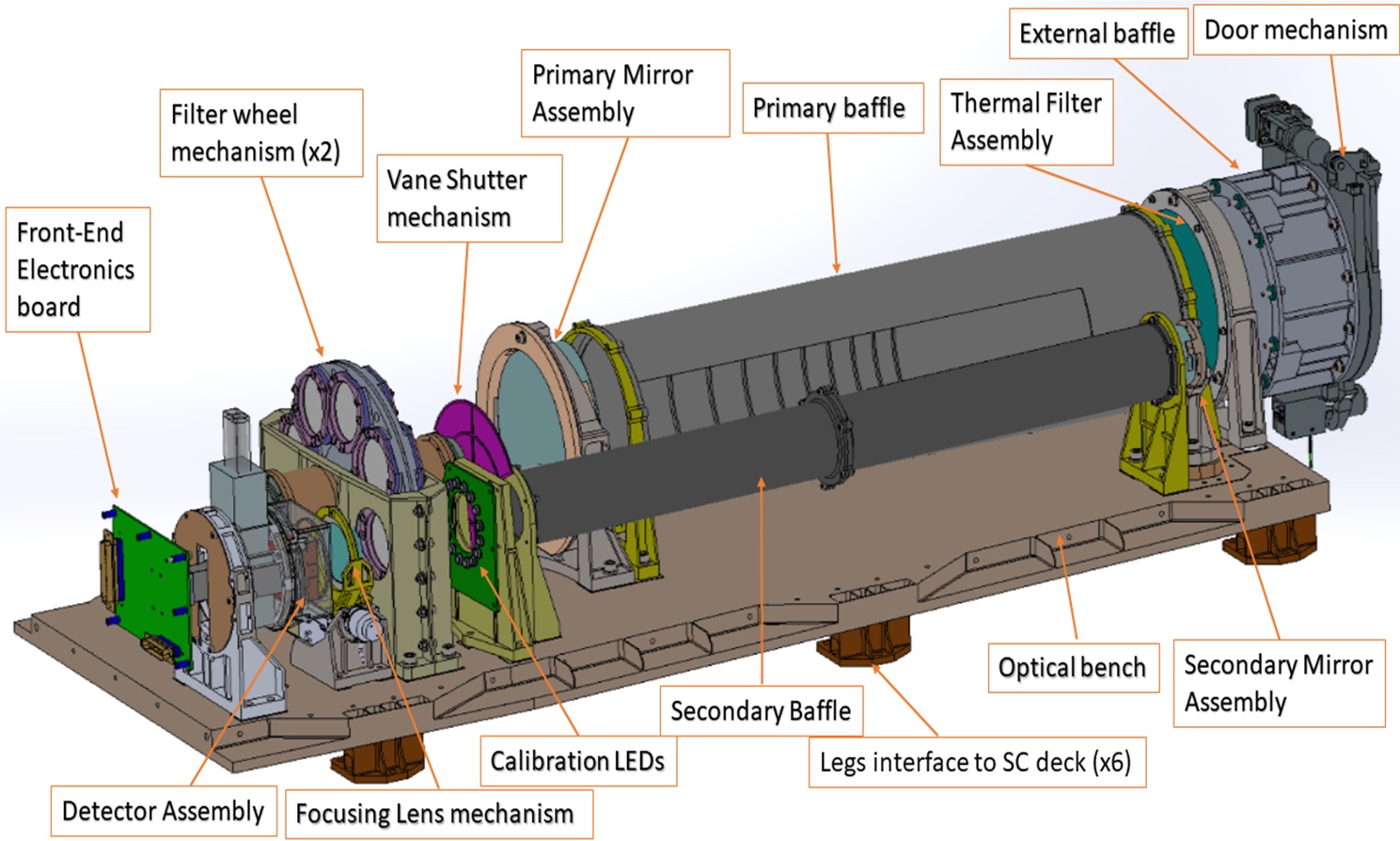} 
 \caption{Internal assembly view of the Solar Ultraviolet Imaging Telescope (SUIT).} \label{fig1}
\end{center}
\end{figure}

SUIT is a Ritchey–Chr\'etien telescope with folded optics carrying a primary mirror with an aperture size of 140~mm and a secondary mirror. It has a set of 11 different science filters centered around different wavelengths between 200{--}400~nm (see Table.~\ref{tab:science_filters}). The images are recorded on a CCD that is kept at the focus. The CCD is UV-thinned and back-illuminated. It has a full well capacity of 190000~e$^{-}$ and a dynamic range of $\sim$12000. SUIT will provide partial disk and full disk images of the Sun with a pixel size of 0.7" and temporal resolution that can range between 4 to 40 seconds, depending on the modes of operation, slicing through the solar atmosphere using various filters. The telescope will produce images with a signal-to-noise (SNR) ratio of 100 in the darkest region of the Sun and a contrast ratio of 10:1 under ambient conditions. 

SUIT has an inbuilt onboard intelligence for tracking regions of interest and detection, localization, and automatic exposure control for flare observations. See \cite{suit, suit1, suit2} for further details on the instrument.

\begin{table}[!h]
\centering
\caption{Filter abbreviations, central wavelength, and bandpass of the SUIT science filters.} \label{tab:science_filters}    
\begin{tabular}{|c|c|c|c|}
\hline
Index & Filter  &		Central		    & 		Bandpass  \\
	  &	name	&		Wavelength (nm)	&		(nm)	\\
\hline
1 & NB1 			& 214.0 		    & 11.0 	\\
2 & NB2 			& 276.7				& 0.4 \\
3 & NB3 			& 279.6 			& 0.4 \\
4 & NB4 			& 280.3				& 0.4 \\
5 & NB5				& 283.2				& 0.4 \\
6 & NB6 			& 300.0 			&1.0 \\
7 & NB7 			& 388.0				&1.0 \\
8 & NB8				& 396.85 			& 0.1 \\
9 & BB1 			& 220.0				& 40.0 \\
10& BB2 			& 277.0 			& 58.0 \\
11& BB3 			& 340.0				& 40.0 \\
\hline
\end{tabular}
\end{table}
\subsection{Visible Emission Line Coronagraph (VELC)}
The Visible Emission Line Coronagraph \citep[VELC;][]{velc} is an internally occulted coronagraph on board Aditya-L1. It can perform simultaneous imaging, spectroscopy, and spectro-polarimetric observations in the lower corona. VELC will image the corona in white light from 1.05{--}3.0~R$_{\odot}$ with a plate scale of 2.5~arcsec/pixel. Spectroscopic and spectro-polarimetric observations will be performed using a multi-slit spectrograph in three emission lines formed at 530.3, 789.2, and 1074.7~nm, with a spectral resolution of 28, 31, and 202~m{\AA}/pixel. The imaging plate scale for 530.3~nm and 789.2~nm, with place scale of 1.25~arcsec/pixel and that for 1074.7~nm is 4.8~arcsec/pixel. The field of view (FOV) for the spectroscopic and spectro-polarimetric channels will be from 1.05{--}1.5~R$_{\odot}$. While the imaging observations in white-light will provide crucial information on the early evolution of CMEs and its kinematics, the spectroscopic observations will help us measure the plasma parameters such as line intensity, electron density, electron temperature and thermal and non-thermal spectral line widths in the ejects. Moreover, the spectroscopic measurements of plasma parameters on quiescent structures in the solar corona will provide crucial information for modeling coronal structures and their heating. The spectro-polarimetric measurements in Fe~\rm{XIII}~1074.7~nm infrared line will provide the topology of the coronal magnetic field for the first time from space.

\begin{figure}[h]
\begin{center}
 \includegraphics[width=0.6\textwidth]{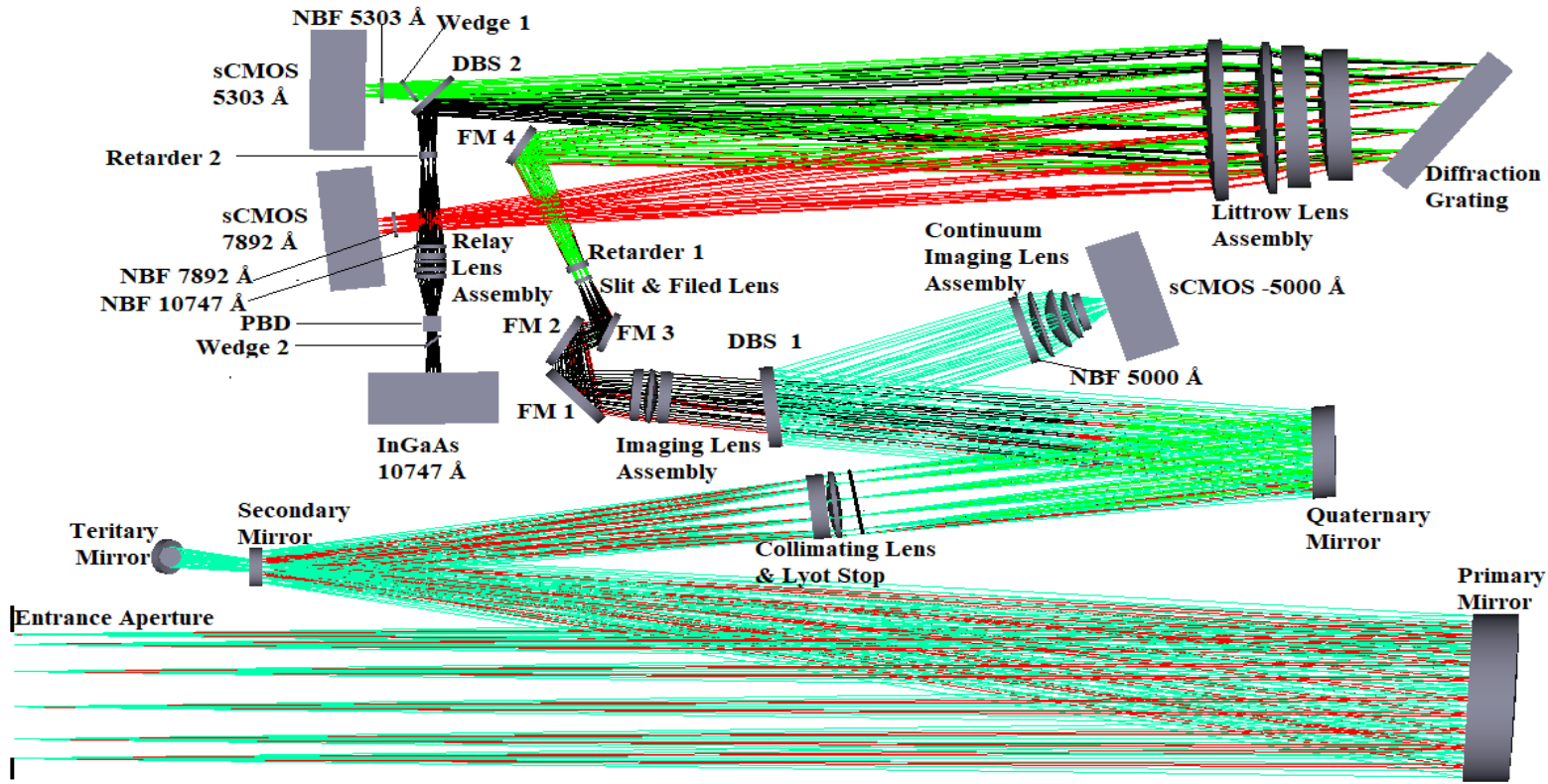} 
 \caption{Optical layout of the Visible Emission Line Coronagraph.} \label{velc}
\end{center}
\end{figure}

The primary science questions that VELC will address are:
\begin{itemize}
    \item What is the most dominant mechanism for coronal heating? What role do MHD waves play?
\item What is the kinematics of CMEs during the early phase of their evolution?
    \item To what degree do coronal inhomogeneities affect the heating and acceleration processes?
    \item What is the magnetic field configuration in the lower corona?
\end{itemize}

VELC has an entrance aperture of 148 mm and an off-axis parabolic primary mirror of 200~mm diameter (see Fig.~\ref{velc}). The light from the primary mirror is then reflected to the secondary mirror, that act as an internal occulter. It allows the disk light to pass through and reflects only the coronal light that is further used for imaging and spectroscopy. The disk light is reflected out into space using another mirror. For imaging and spectroscopy in visible, VELC uses three sCMOS detectors. However, for spectro-polarimetry using Fe~\rm{XIII} in infrared, VELC uses InGaAs detector system. See \cite{velc, velc1, velc2} for further details on the instrument. Like SUIT, VELC also developed an onboard algorithm for automatic CME detection \cite{Patel}.

\subsection{X-Ray Observations of the Sun with Aditya-L1}
The Aditya-L1 mission has two full-Sun integrated X-ray spectrometers \citep{solex_hel1os}, {\it viz.} the Solar Low Energy X-ray Spectrometer (SoLEXS) and High Energy L1 Orbiting X-ray Spectrometer (HEL1OS) with their primary goal of studying solar flares. 

\begin{figure}[h]
\begin{center}
 \includegraphics[width=0.4\textwidth]{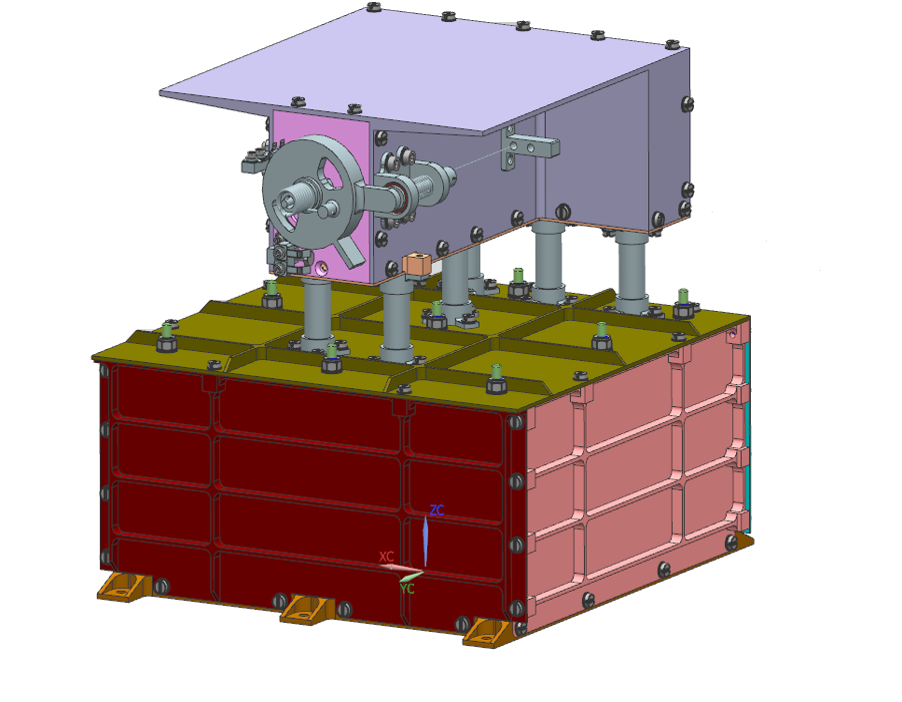} 
  \includegraphics[width=0.4\textwidth]{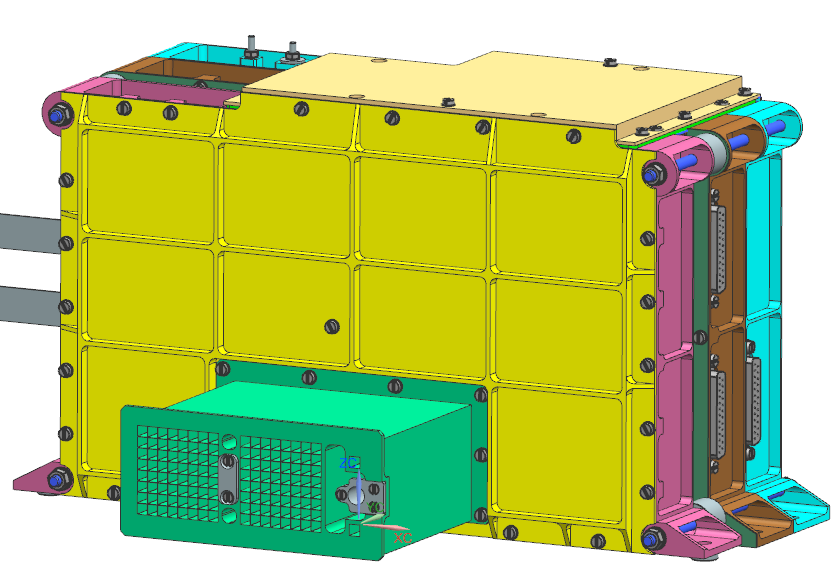} 
 \caption{Concept design of SoLEXS (Left panel) and HEL1OS (right panel).} \label{solexs}
\end{center}
\end{figure}

SoLEXS (see left panel of Fig.~\ref{solexs}) covers the energy band from 1 to 30~keV with a spectral resolution of $<$250~eV at 6~keV and primarily covers the thermal energy part of the flares. SoLEXS carries two identical detectors but with different aperture sizes to observe smaller as well as larger flares. The detector with the larger aperture will be sensitive to smaller flares, whereas that with the smaller aperture will be sensitive to larger flares. It will have onboard processing to help provide spectra at a cadence of 1s during flares. Moreover, it will have onboard intelligence to detect the start of flares. This input is also provided to the SUIT payload for their algorithm to localize and start the flare mode operation.

HEL1OS (see right panel of Fig.~\ref{solexs}) covers the energy band from 10 to 150~keV, with the prime aim of studying the impulsive phase of the flares, where most of the emission is shown to be non-thermal. Therefore, it will provide unique observations to help identify the non-thermal energy released during flares. HEL1OS will carry two different types of detectors, {\it CZT and CdTe}. There are two CZT detectors that will cover the energy range of 20 to 150~keV. The CdTe detector is sensitive from 10 to 40~keV and will be used for detailed spectroscopic studies. For further details of SoLEXS and HEL1OS, see \cite{solex_hel1os}.

The two spectrometers together will cover a broad range of energy, i.e., from 1~keV to 150~keV, with excellent spectral resolution. The primary science objectives of the two spectrometers together are:

\begin{itemize}
    \item explosive energy release, particle acceleration, and transport using high-resolution spectral and imaging information.
    \item trigger mechanism of flares/eruptive processes.
    \item cut-off energy between thermal and non-thermal emission as a function of flare evolution.
    \item SXR, HXR and CME association
\end{itemize}

\subsection{In-situ measurements at L1 with Aditya-L1}
Among the seven payloads, three are {\it in situ} instruments, namely the Aditya Solar Wind Particle Experiment (ASPEX), the Plasma Analyser Package for Aditya (PAPA), and Magnetometer (MAG). While ASPEX and PAPA are particle detectors, the MAG instrument is flux gate magnetometers \citep{aspex_papa_mag}.

\begin{figure}[h]
\begin{center}
 \includegraphics[width=0.6\textwidth]{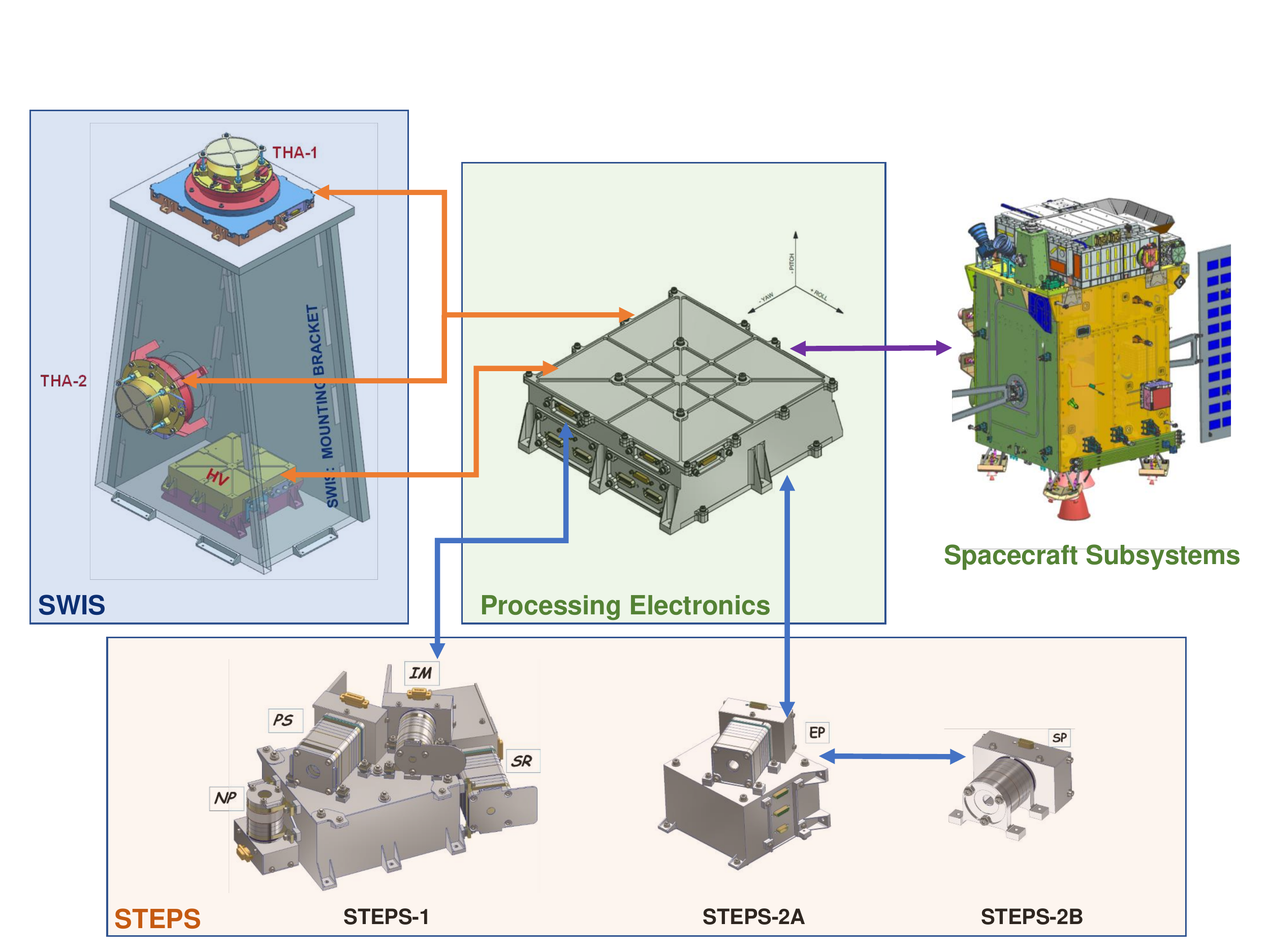} 
 \caption{ASPEX payload and its various subsystems. THA, HV stand for Top Hat Analyzer and High Voltage unit respectively. Note, SWIS consists of THA-1 and THA-2 packages and STEPS consists of STEPS-1 (Sun Radial or SR, Intermediate or IM, Parker Spiral or PS, North Pointing or NP) and STEPS-2 (Earth Pointing or EP and South Pointing or SP) packages.} \label{aspex}
\end{center}
\end{figure}

ASPEX (see Fig.~\ref{aspex}) contains two ion spectrometers \citep[][]{aspex_papa_mag, aspex1, aspex2}, {\it viz.} the Solar Wind Ion Spectrometer (SWIS) and the Supra Thermal and Energetic Particle Spectrometer (STEPS). While SWIS covers the low energy particles in the range 0.1{--}20~keV, STEPS aims to detect high energy particles in the energy range 20~keV/nucleon{--}5~MeV/nucleon. The primary science objectives of ASPEX are to understand better the origin, acceleration, inter-relationship and an-isotropy of ions (primarily protons and alphas) in the background solar wind during the passage of transient events such as CMEs and co-rotating interaction regions (CIRs). By measuring (protons and alphas separated in some directions and integrated in other directions) these ions, in the slow and fast solar wind (360 deg along and across the ecliptic) as well as in the suprathermal and solar energetic particle (SEP) domains in multiple directions (Sun Radial, Intermediate, Parker Spiral, North Pointing, South Pointing and Earth Pointing), ASPEX will be able to investigate the characteristics of CMEs, SIRs (or CIRs) and energetic ions. Further, based on the direction-resolved observations of these ions, ASPEX may pave the way for the efficient prediction capability for the arrival of these transients at the L1 point.

\begin{figure}[h]
\begin{center}
 \includegraphics[width=0.6\textwidth]{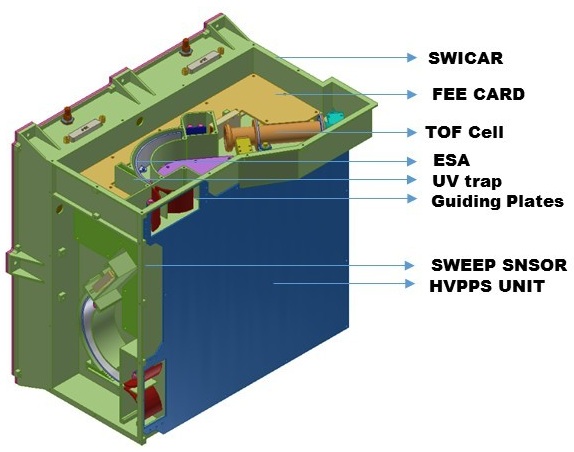} 
 \caption{Schematics of the PAPA payload. Different components are marked.} \label{papa}
\end{center}
\end{figure}

The Plasma Analyzer Package for Aditya (PAPA) aims at studying the composition of solar wind and its energy distribution (in the range from 0.01{--}3~keV for electrons and 0.01{--}25~keV for ions) continuously throughout the mission’s lifetime. PAPA contains two sensors (see Fig.~\ref{papa}); Solar Wind Electron Energy Probe (SWEEP) to measure the solar wind electron flux and Solar Wind Ion Composition AnalyseR (SWICAR) to measure the ion flux and composition as a function of direction and energy. SWEEP will measure only electron parameters whereas SWICAR has two modes of operation – ion mode where ion parameters are measured and electron mode where electron parameters are measured. These two modes in SWICAR are mutually exclusive. Data from PAPA would provide detailed knowledge of the solar wind condition with high time resolution. SWICAR would also provide: (1) the elemental composition of solar wind low energy ions, and (2) the differential energy spectra, abundances of dominant ion species in the mass range 1-60 amu with energies between 0.01{--}25 keV/q. From the differential energy spectra, the 'key parameters' such as bulk speed, density and kinetic temperature of the dominant solar wind ion species could be regularly derived; inferences could be made on the coronal temperatures; plasma sources of suprathermal ion populations could be identified, and the nature and dynamics of solar wind plasma could be quantified.

The MAG instrument package consists of two tri-axial flux gate magnetometers (FGMs). The two FGMs will be mounted on a 6-meter-long boom. The two FGMs are used because the spacecraft is not magnetically clean. Therefore, the presence of two identical FGMs is used to obtain the differential interplanetary magnetic field (IMF) by nullifying the effect of the magnetic field from the spacecraft. The primary science objective of MAG is to measure and monitor the magnitude and direction of the IMF locally at the L1 point. The magnitude and direction of IMF associated with CMEs are two crucial parameters for space weather forecasting. The other important science objective of the MAG payload is to observe the CMEs from the Sun coming towards the Earth at the L1 point \citep[][]{mag1}.

Apart from the above primary objectives, the secondary objectives of this magnetic field experiment are (1) Near Earth Space Weather studies thereby combining the in-situ magnetic field measurements onboard Aditya-L1 spacecraft with ground based measurements of geomagnetic field; and (2) the detection of plasma wave signatures at the L1 point with the magnetic field and solar wind particle velocity measurements. These plasma (possibly Alfv\'en) wave signatures are detected in the magnetic field and solar wind observations onboard ACE and WIND \citep[][]{mag2}.

The measurements from ASPEX and PAPA combined with those from MAG will provide crucial insights towards understanding the dynamics of the inner heliosphere and physical processes occurring therein and will help us towards better forecasting of space weather events.
\section {Summary} 
Aditya-L1 is truly a satellite for multi-messenger solar physics. With the suite of instruments on board, Aditya-L1 will perform remote sensing covering a broad range of the electromagnetic spectrum and in situ measurements in a wide energy range from the same platform. For the first time, it will provide regular measurements of the coronal magnetic field from space and spatially resolved solar spectral irradiance in the near ultraviolet. Moreover, Aditya-L1 will observe flares in both soft X-ray and hard X-ray from the same platform. Similarly, the two particle instrument, along with the magnetometer, will be highly beneficial for the dynamics of the inner heliosphere and space weather studies.

\begin{acknowledgments}
The generous travel support from the IAU is gratefully acknowledged. Aditya-L1 is a mission of the Indian Space Research Organization (ISRO). We acknowledge ISRO's all round support in developing the payloads and proving this mission of opportunity.
\end{acknowledgments}
\def\planss{{Planetary and Space Science}}
\def\grl{{GRL}}
\def\solphys{{Solar Physics}}
\def\ssr{{Space Science Reviews}}    
\def\apj{{ApJ}}    
\def\apjs{{ApJS}}    
\def\nat{{Nature}}    
\def\jgr{{JGR}}    
\def\apjl{{ApJ Letters}}    
\def\aap{{A\&A}}   
\def\mnras{{MNRAS}}
\def\aj{{AJ}}
\let\mnrasl=\mnras
\def\hia{{Highlights of Astronomy}}
\bibliographystyle{aa}
\bibliography{biblio.IAU372}
\end{document}